\documentclass[]{cupconf} 
\usepackage{graphicx}

\title[The N/O evolution on galaxies]
     {The N/O evolution on galaxies:\\
       the role played by the star formation history}
      
\author[M.~Moll\'{a} et al.]{\\ Mercedes Moll\'{a}$^1$ 
\\ Jos\'{e} M. V\'{\i}lchez$^2$ 
\\ Angeles I. D\'{\i}az$^{3}$
\\ Marta Gavil\'{a}n$^3$}
\affiliation{ $^{1}$ Departamento de Investigaci\'{o}n B\'{a}sica, CIEMAT,
 Avda. Complutense 22, 28040, Madrid, (Spain) \\
e-mail:{mercedes.molla@ciemat.es}\\
$^{2}$ Instituto de Astrof\'{\i}sica de Andaluc\'{\i}a (CSIC), 
Apdo. 3004, 18080 Granada, (Spain) \\
e-mail:{jvm@iaa.es}\\
$^{3}$ Departamento de F\'{\i}sica Te\'{o}rica,
Universidad Aut\'onoma de Madrid, 28049 Cantoblanco, Madrid 
(Spain)\\
e-mail:{angeles.diaz@uam.es} \\
e-mail:{marta.gavilan@uam.es}
}

\begin{document}
\maketitle

\begin{abstract}
We study the evolution of nitrogen resulting from a set of spiral and
irregular galaxy models computed for a large number of input mass
radial distributions and with various star formation efficiencies.  We
show that our models produce a nitrogen abundance evolution in good
agreement with the observational data. In particular, low N/O values
for high-redshift objects, such as those obtained for Damped Lyman
Alpha galaxies can be obtained with our models simultaneously to
higher and constant values of N/O as those observed for irregular and
dwarf galaxies, at the same low oxygen abundances $\rm 12+log(O/H)
\sim 7$ dex.  The differences in the star formation histories of the
regions and galaxies modeled are essential to reproduce the
observational data in the N/O-O/H plane.
\end{abstract}
\newpage
\section{The plane N/O {\sl vs} O/H}

\begin{figure}
\begin{center}
\includegraphics[width=\textwidth,angle=0]{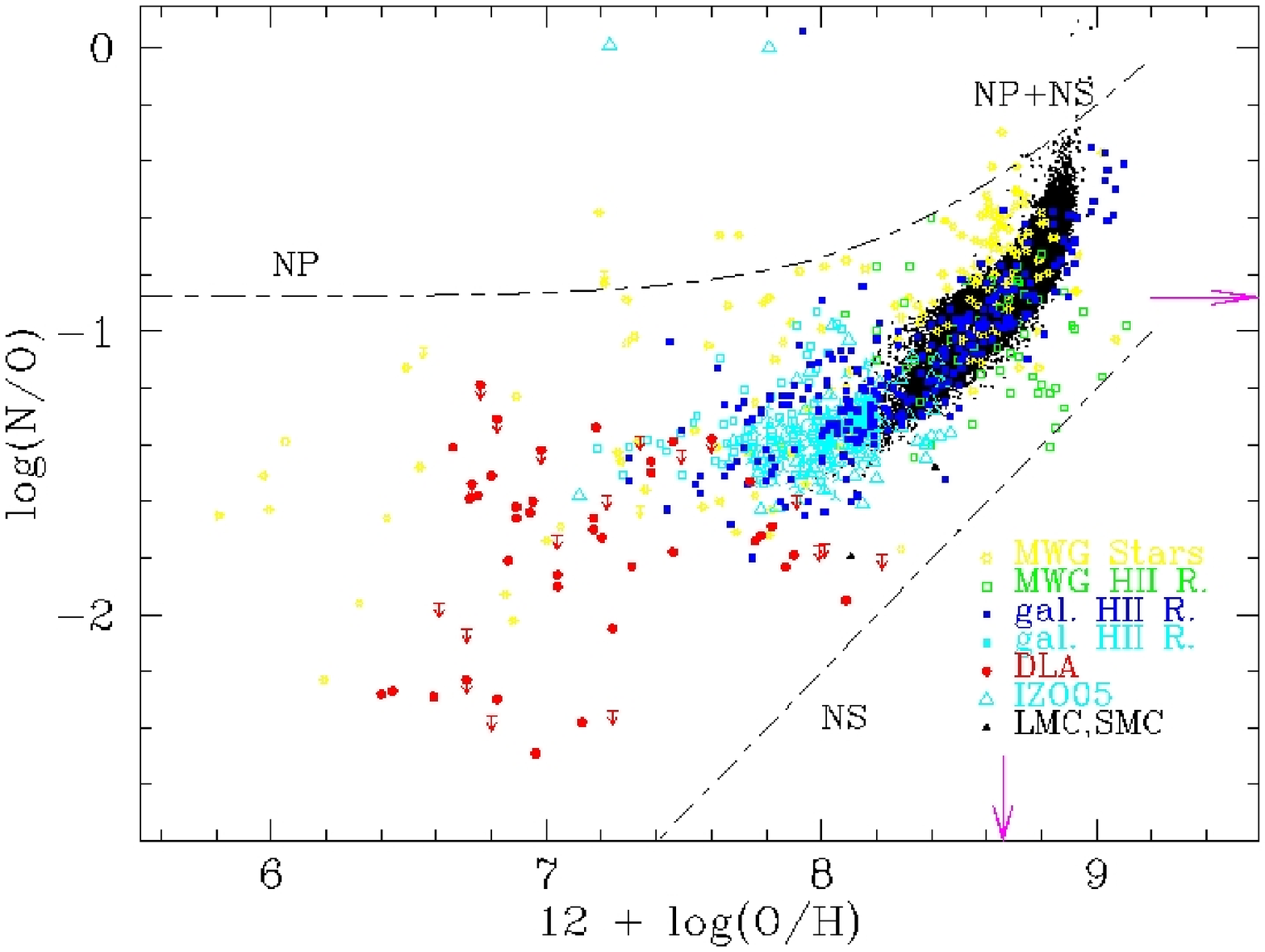} 
\caption{The relative abundance $log(N/O)$ {\sl vs} the oxygen
abundance $12 + log(O/H)$ observations. Black small points are the
SDSS galaxy data from \cite{lian06}. Yellow open dots are data for
Galactic stars.  The green open, solid blue and cyan squares are the
H{\sc ii} region data for MWG, metal-rich and metal-poor galaxies,
respectively, (including in the last ones data from \cite{nava06} and
\cite{izo99,izo06} estimates). The large cyan open triangle is lowest
galaxy known data, given by \cite{izo05}. Red fill dots are values for
Damped Lyman Alpha objects.}
\end{center}
\label{data}
\end{figure}

The whole set of data in the plane N/O-O/H (for references od
data see Table 1 from \cite{mol06}), such as they
are shown in Fig.~\ref{data}, are limited by three possible  
theoretical lines:

\begin{enumerate}
\item the line defined by the secondary nitrogen behavior, NS, when N 
needs a seed of O to be created.
\item the corresponding one for primary nitrogen, NP, when N is created 
directly from H or He  for any oxygen abundance
\item NS+NP when both contributions there exist
\end{enumerate}

It is evident that a primary N contribution must exist.  Actually, a
secondary production of N is expected for most stars as corresponds to
a CNO cycle, but it is well established that a primary production
should arise from intermediate mass stars ($\rm 4 \leq M < 8
M_{\odot}$) that suffer dredge-up and hot-bottom burning episodes
during the asymptotic giant branch (\cite{rv81,hoe97}).

\section{The evolution of nitrogen in the Milky Way Galaxy}
\begin{figure*}
\begin{center}
\includegraphics[height=0.45\textheight,angle=0]{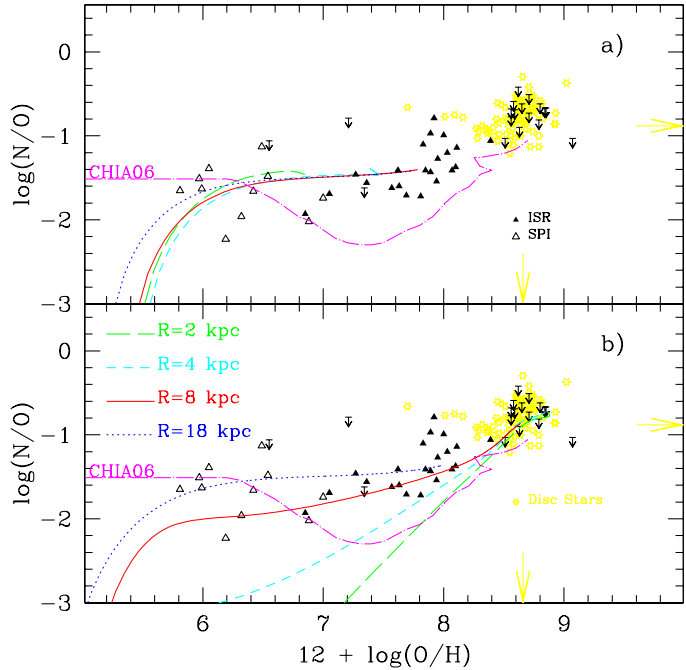} \hfill
\caption{The evolution of the relative abundance $log(N/O)$ {\sl vs}
the oxygen abundance $12 + log(O/H)$  for four regions of MWG 
located at galactocentric distances $R=2$,4, 8 and 18
kpc as labelled. Upper panel for the halo regions. Bottom panel for the disc
regions. The data correspond to Galactic stars of the halo (black
triangles) and of the disc (yellow open dots).}
\label{mwg}
\end{center}
\end{figure*}
Using massive star yields from \cite{woo95} and low and intermediate
star yields from \cite{gav05} with a certain proportion of primary N,
dependent on metallicity, a Galactic chemical evolution model (GCEM)
has been calculated (\cite{gav06}) which reproduces most data of the
Milky Way Galaxy (MWG).  In Fig.~\ref{mwg} the MWG model results are
shown separately for the halo (a) and for the disk (b). In each panel
the evolution for four regions, located at different distance of the
galaxy center are represented.  The evolution in the halo is similar
for all regions, reaching the level observed of relative abundance N/O
and maintaining an almost constant ratio.  This behavior is in
agreement with the observed trend for the halo stars, in particular
the recent ones from \cite{isr04} and from \cite{spi05}. In the disk,
however, each radial region has its own evolution, as corresponds to
the different star formation history that occurs in each one. The
outer regions with quiet star formation histories have still low
oxygen abundances when the intermediate stars begin to eject their
primary nitrogen, while the inner regions, that suffer a strong and
early star formation rate, have created many generations of stars, ans
consequently produced high oxygen abundances before this NP appears in
the interstellar medium. For $R=2$ or 4 kpc this NP appears at almost
solar oxygen abundances, shown in the figure like a smoother increase
or even as a flattening of the evolutionary track.  For comparison
purposes we also draw the Solar region model from \cite{chia06} where
NP proceeds from low Z massive stars rotating at very high
velocity. Our model reproduces better the observations.

\section{The results of our simulations for all time steps}
\begin{figure}
\begin{center}
\includegraphics[width=0.75\textwidth,angle=0]{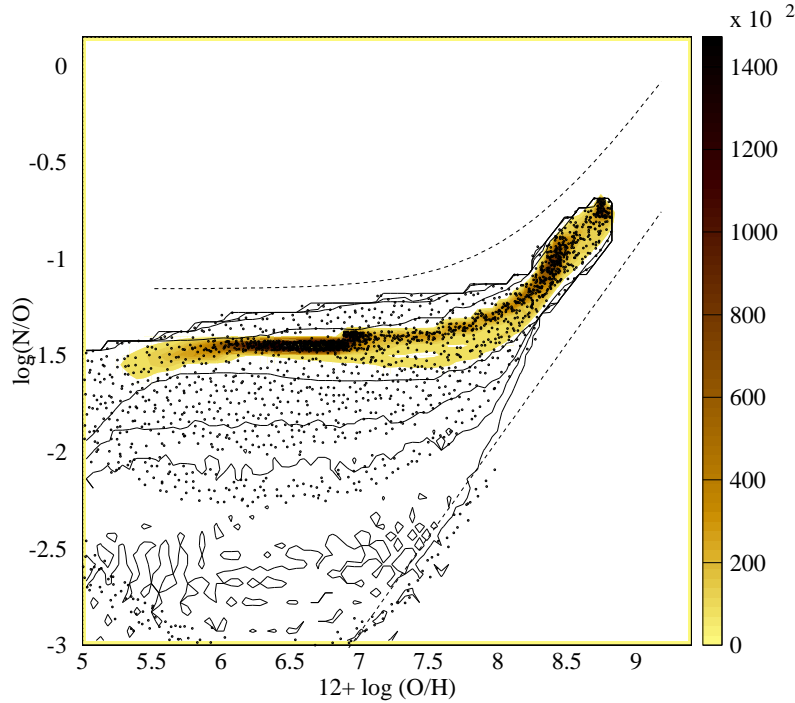} \hfill
\caption{The relation between N/O and O/H for the simulated
galaxies. The complete results for all modeled regions shown as small
dots. Colored contours represent the isolines of number of points
included within them (see relative scale to the right), while solid
lines represent the regions with 0.3, 0.03, 0.012 and 0.003 \%,
respectively, of the total number of points.  The data region is
limited by the two dashed lines.}
\label{all}
\end{center}
\end{figure}
Since these stellar yields seem to be good enough to fit the Galactic
data we apply them to a wide grid of theoretical galaxies with
different total mass (\cite{mol05}).  Using different star formation
rate efficiencies, we obtain different results of abundances for the
whole set of models. For details, see \cite{mol06}. In Fig.~\ref{all}
we show the results, for all calculated times and models, as small
dots. Over them we have plot some contours in yellow
levels. Furthermore, we draw some other contours as solid lines for
regions with 0.3, 0.03, 0.012 and 0.003 \%, respectively, of the total
number of points.  The model results for all time step is also in
agreement with the observed trends showing:

\begin{itemize}
\item The low mass galaxies maintain a flat behavior in the plane
N/O-O/H with an almost constant abundance N/O even for oxygen
abundances as low as $12+log(O/H) \sim 6.5-7$
\item The massive galaxies show a different trend in that plane
depending on their star formation history. The most efficient in
forming stars, that is those that create them by mean of a high and
early star formation rate, seem to have a secondary behavior from the
first moment with a very steep evolution of N/O {\sl vs} O/H.
\item This behavior behaves more like {\sl primary+secondary}, that is
with a smaller slope in that graph, when stars form less efficiently
reaching the primary line of low mass galaxies when the star formation
rate is very low. These simulated galaxies would appear as low
brightness massive galaxies.
\item The predicted dispersion is also large, as shown by the observations.
\end{itemize}

\section{The present time abundances for a grid of models}

\begin{figure}
\begin{center}
\includegraphics[width=\textwidth,angle=0]{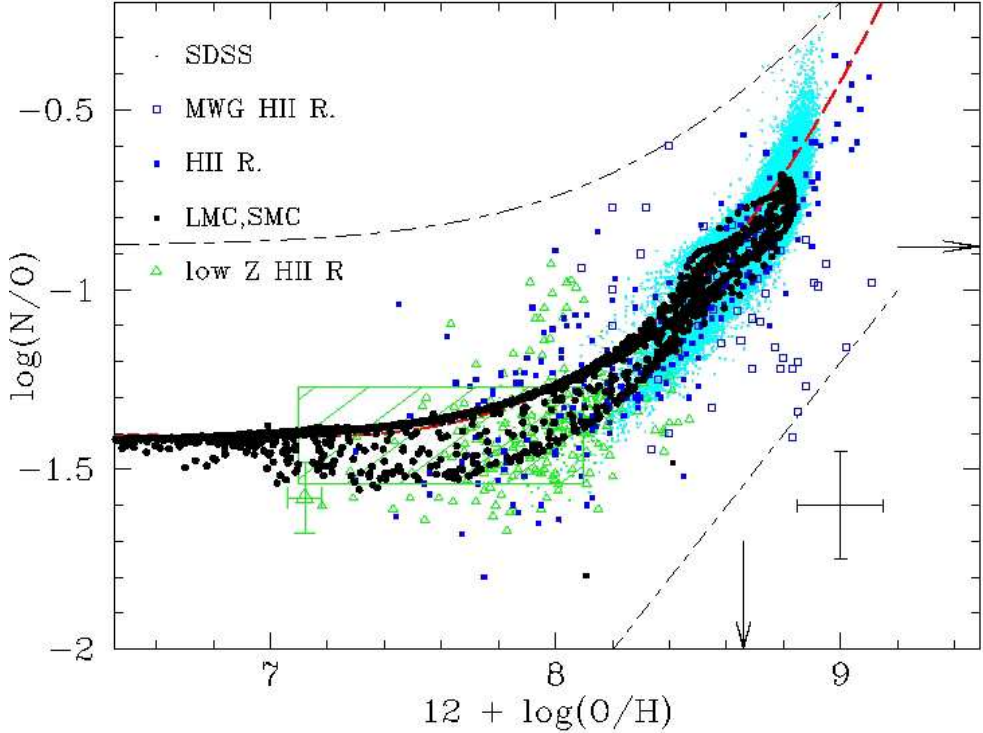}
\caption{The relative abundance $log(N/O)$ {\sl vs} the oxygen
abundance $12 + log(O/H)$ resulting from our models for the present
time as full black dots compared with the data corresponding to
Galactic and extragalactic H{\sc ii} regions of SDSS 
galaxies (\cite{lian06}, small cyan and green points for metal-rich
and metal-poor galaxies), to galactic and extragalactic H{\sc ii} regions
from authors of Table~1 of \cite{mol06} (blue open and full squares,
respectively), and for low metallicity galaxies
(\cite{nava06,izo99,izo06} as green triangles).  The long-dashed line
is the least squares fitting function to models. The large triangle
around $12+log(O/H) \sim 7.1$ is the value found by \cite{izo05} for
the lowest-metallicity star-forming galaxy known.}
\label{present}
\end{center}
\end{figure}

We represent the present time model results in the plane N/O vs O/H in
Fig.~\ref{present} as small black points, that we compared with the
galactic and extragalactic H{\sc ii} regions data. The models are fitted by a
least-squares polynomial function:
\begin{eqnarray*}
log(N/O)& = &-1149.31+1115.23x-438.87x^{2}+90.05x^{3}\\
        &   & -10.20x^{4}+ 0.61x^{5} -0.015x^{6}
\end{eqnarray*}
where $x=12+log(O/H)$.

We demonstrate that using yields from LIM mass stars and a model with
star formation efficiencies variable for each galaxy, we may obtain
points in the plane N/O-O/H in agreement with data, even for what
refers to the observed dispersion which also may partially be
simulated with our models. In the figure we have plot a shaded
rectangle showing the region where abundances N/O fall within a
Gaussian distribution, which implies that differences shown by points
in this zone with models may be mainly imputed to observational errors
(\cite{nava06}).

\section{The time or redshift evolution of N/O}

\begin{figure}
\includegraphics[width=0.75\textwidth,angle=-90]{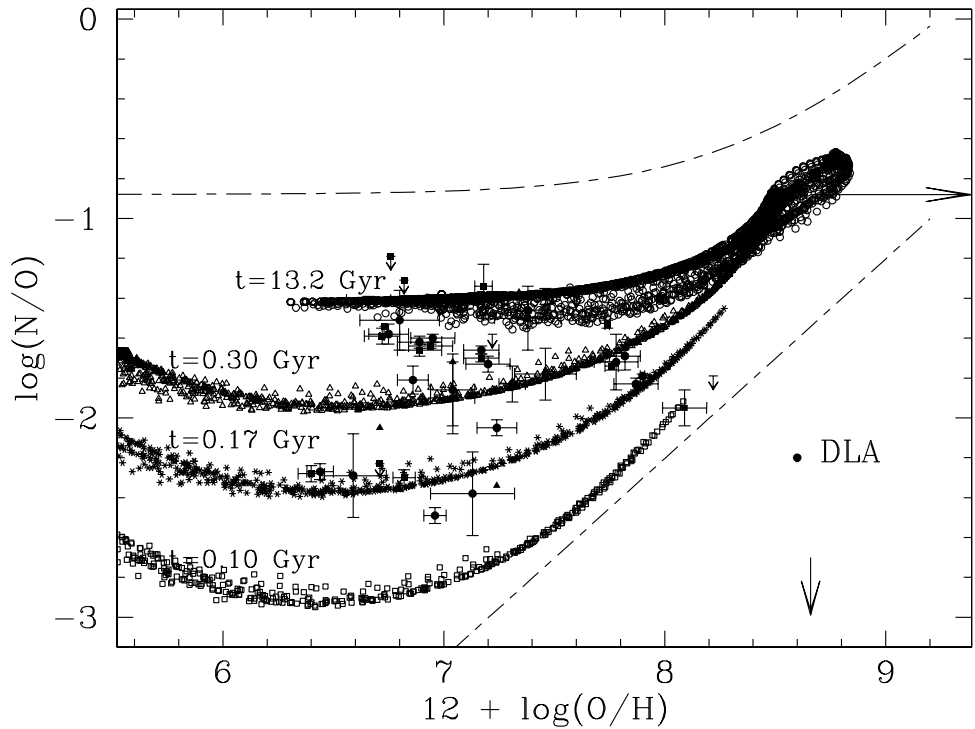}
\caption{The relation N/O -- O/H for different evolutionary times:
 $\rm t=0.1$, $\rm t=0.17$, $\rm t=0.30$, and $\rm t=13.2$ Gyr, as
marked in the figure. Data as full dots or triangle correspond to DLA
objects.}
\label{redshift}
\end{figure}

Finally, in order to explore the time evolution of the modeled
abundances, we show in Fig.~\ref{redshift} the results for four
different time steps: $\rm t=0.1$, $\rm t=0.17$, $\rm t=0.30$, and
$\rm t=13.2$ Gyr.  These epochs would correspond to redshifts
z$\sim$3.8, 3.7, 3.5, and 0, respectively, for a cosmology with
H$_{0}=71$, $\Omega_{M}=0.30$, $\Omega_{\Lambda}=0.70$, if the
formation of these spiral and irregular galaxies occurred at a
redshift z$\sim 4$. A large gap appears between the results for each
one of our time steps. In particular, our models predict a feature
similar to the so-called {\sl second plateau} which Centuri\'{o}n et
al. (2003) claim to exist at a metallicity around $\rm 12+log(O/H)\sim
7$, which appears at a level of $\rm \log{(N/O)}\sim -2.3$ for $\rm
12+log(O/H) \sim 6.5-7$, while most points appear at a higher level of
N/O ($ \ge -1.8$) for a similarly low O/H abundance. In fact, no gap
is apparent between models at 1.1 and 13.2 Gyr, thus making it
difficult to discriminate objects at a redshift up to $z < 2.5$ from
those at redshift z=0 in the N/O-O/H plane, as it is actually the
case.  The abundances predicted for galaxies at redshift $z \ge 3.$
are far enough from the rest of the points in the plot so as to
disentangle them from a given data sample.

\section{Concluding remarks}

These results are only partially due to the dependence with
metallicity of the yields we use. This contribution of primary N,
larger for low metallicity stars than for high metallicity ones, is
important since it allows to us to obtain tracks in the plane N/O {\sl
vs} O/H flatter than the ones predicted with a constant contribution
of the primary N. We must to do clear, however, that if we eliminate
the metallicity dependence from the yields, we still obtain different
tracks for regions with different star formation histories, such as
we demonstrate with detail \cite{mol06}.

The absolute level of observations and the fine tuning of the observed
shape in the plane N/O-O/H for the present time data, however, is only
reproduced if the stellar yields, in turn, give the right level of
primary nitrogen and have the adequate dependence on metallicity. The
chemical evolution models must also be well calibrated in order to
predict star formation histories able to produce abundances in
agreement with data. Therefore, the adequate selection of metallicity
dependent LIM star yields joined to the accurate chemical evolution
models are able to predict abundances for N and O which reproduce the
complete set of data in the plane N/O {\sl vs} O/H.

The main outcomes of this work are the following:
\begin{itemize}
\item The evolutionary track of a region or galaxy in the plane N/O-O/H
is very dependent on the star formation history. When this occurs as
a strong burst the evolution follows a {\sl secondary} behavior
while a continuous, quiet and low star formation rate gives a flat
slope in that plane. 
\item Intermediate star formation histories are between
the two referenced extreme trends, producing this way a large dispersion of the
corresponding data which is similar to the observed one
\item The final points of our realizations, corresponding to the present time
abundances, reproduce very well the trend given by H{\sc ii} regions data.
\item The trend given by low-metallicity data from the MWG halo is
 also well fitted.
\item The data proceeding from high-redshift objects may be easily explained
with  the abundances calculated for other evolutionary times. In particular
something  like a {\sl second plateau} appears when results for $z> 3$ are 
plotted.

\end{itemize}


\begin{thebibliography}{99}
	 
\bibitem[{Chiappini et al.}{(2006)}]{chia06} 
Chiappini C., Hirschi R., Matteucci F., Meynet G., Ekstrom S., Maeder
A., 2006, astro, arXiv:astro-ph/0609410

\bibitem[{Gavil{\' a}n}, {Buell} \&  {Moll\'{a}}{ (2005)}]{gav05} 
{Gavil{\' a}n} M., {Buell} J.~F., {Moll\'{a}} M., 2005, \textit{A\&A}, 432, 861

\bibitem[{Gavil{\' a}n}, {Moll{\' a}} \&  {Buell}{ 2006}]{gav06} 
{Gavil{\' a}n} M.,  {Moll{\' a}} M., {Buell} J.~F., 2006, 
\textit{A\&A}, 450, 509

\bibitem[Israelian et~al. (2004)]{isr04}
{Israelian} G.,  {Ecuvillon} A.,  {Rebolo} R.,  {Garc{\'{\i}}a-L{\' o}pez} R.,
  {Bonifacio} P.,    {Molaro} P.,  2004, \textit{A\&A}, 421, 649

\bibitem[{Izotov et al.}{ 2006}]{izo06} 
Izotov Y.~I., Stasi{\'n}ska G., Meynet G., Guseva N.~G., Thuan T.~X., 2006, 
A\&A, 448, 955 

\bibitem[{{Izotov}, {Thuan} \& {Guseva}}{(2005)}]{izo05} 
{Izotov} Y.~I., {Thuan}  T.~X., {Guseva} N.~G., 2005, \textit{ApJ},2,2

\bibitem[{Izotov \& Thuan}{ 1999}]{izo99} 
Izotov Y.~I., Thuan T.~X., 1999, ApJ, 
511, 639

\bibitem[{Liang et al.}{(2006)}]{lian06} 
Liang Y.~C., Yin S.~Y., Hammer F., Deng L.~C., Flores H., Zhang B., 2006, 
ApJ, 652, 257 

\bibitem[{{Moll{\' a}} \&  {D{\'{\i}}az}}{ 2005}]{mol05}
{Moll{\' a}} M.,  {D{\'{\i}}az} A.~I.,  2005, \textit{MNRAS}, 358, 521

\bibitem[{Moll{\'a} et al.}{(2006)}]{mol06}
Moll{\'a} M., V{\'{\i}}lchez J.~M., Gavil{\'a}n M., D{\'{\i}}az A.~I.,
2006, MNRAS, 372, 1069

\bibitem[{Nava et al.}{ 2006}]{nava06} 
Nava A., Casebeer D., Henry R.~B.~C., Jevremovic D., 2006, ApJ, 645, 1076 

\bibitem[{Renzini} \& {Voli}{ 1981}]{rv81} 
{Renzini} A.,  {Voli} M.,  1981, A\&A, 94, 175

\bibitem[{{Spite} et~al.}{ (2005)}]{spi05}
{Spite} M., {Cayrel} R., {Plez} B., {Hill} V., {Spite} F., {Depagne}
  E., {Fran{\c c}ois} P., {Bonifacio} P., {Barbuy} B., {Beers} T.,
  {Andersen} J., {Molaro} P., {Nordstr{\" o}m} B., {Primas} F., 2005,
  \textit{A\&A}, 430, 655

\bibitem[{{van den Hoek} \& {Groenewegen}}{ 1997}]{hoe97}
{van den Hoek}, L.~B. \& {Groenewegen}, M.~A.~T. 1997, A\&AS, 123, 305

\bibitem[{{Woosley} \&  {Weaver}}{ (1995)}]{woo95}
{Woosley} S.~E.,  {Weaver} T.~A.,  1995, \textit{ApJS}, 101, 181

\end{thebibliography}
\end{document}